\documentclass[dvips]{article}

\usepackage{icrc2011}

\title{Search for VHE signals from microquarsars with MAGIC}

\newcommand{\etal}{\MakeLowercase{\textit{et al. }}} 
\shorttitle{Zanin, R. \etal microquasars observations with MAGIC}

\authors{Zanin, R.$^{1}$, Saito, T.$^{2}$,Zabalza, V.$^{3}$, Bordas, P.$^4$, Jogler,
  T.$^2$, Cortina, J.$^{1}$, Paredes, J.~M.$^3$, Ribo, M.$^3$, Rico,
  J.$^1$ for the MAGIC collaboration}
\afiliations{$^1$IFAE Barcelona, Spain\\
$^2$MPI-Munich, Germany\\
$^3$Departament d'Astronomia i Meteorologia, Institut de Ci\'{e}ncies del Cosmos
(ICC), Universitat de Barcelona (UB-IEEC), Spain\\
$^{4}$ INTEGRAL Science Data Centre, Universit\'{e} de Gen\'{e}ve\\
$^{5}$ Instituci\'{o} Catalana de Recerca y Estudis Avançats (ICREA), Barcelona, Spain}
\email{roberta@ifae.es}

\abstract{
Microquasars are accreting binary systems displaying relativistic
radio jets where very high energy gamma rays might be produced via
Inverse Compton scattering. The detection of the microquasar Cygnus
X-3 above 100 MeV by both Fermi/LAT and AGILE satellites, together
with the short flares from Cygnus X-1 reported by AGILE, confirmed that
this class of astrophysical objects are interesting candidates for
very high energy gamma-ray observations. The stand-alone imaging
atmospheric Cherenkov MAGIC telescope, and later on the recently
inaugurated stereoscopic system, made a significant effort to search
for signals from microquasars. This paper reviews all MAGIC results
of Cygnus X-3, Cygnus X-1, Scorpius X-1, and SS 433 observations. The
stand-alone MAGIC telescope observed Cygnus X-3 for almost 60 hrs from
March 2006 until August 2009 in many different X-ray spectral states
where a very high energy emission is predicted and also simultaneously
with a flux enhancement at high energies detected by Fermi/LAT. No
significant signal was found in any of the observed conditions. The
MAGIC stereoscopic system pointed at the Z-type low-mass X-ray binary
Scorpius X-1 in May 2010 for ~8 hrs. Simultaneous soft X-ray
measurements allowed to define the X-ray spectral state of the source
which could emit very-high-energy photons in the horizontal
branch. MAGIC did not detect the source and put some constraints on
the maximum TeV luminosity to jet power ratio. Further observations of
Cygnus X-1 were also carried out with the new and twice as sensitive
stereoscopic system in Autumn 2009 for a total amount of 30
hrs. Finally, we also report on SS 433 observations obtained during
two periods of minimum absorption processess from the accretion disk.
}


\begin{document}
\maketitle

\section{Introduction}
Microquasars are X-ray binaries which display collimated relativistic
jets at radio frequencies \cite{Mirabel1999}. They consist of an
accreting compact object, either a black hole or a neutron star, and a
companion star. Depending on the mass of the companion, X-ray binaries
are classified as high-mass (HMXBs) and low-mass (LMXBs) X-ray binaries. 

It has been proposed that particles accelerated in the relativistic
ejections might produce $\gamma$-ray emission via Inverse Compton (IC)
scattering \cite{Atoyan1999}. In the case of HMXBs, the massive
and luminous stellar companion produces an intense target photon field 
for the IC scattering. However, this photon field could also absorb the
produced $\gamma$-rays through pair production making the detection of
the source at very high energies (VHEs) very difficult. This absorption effect is almost
negligible for the LMXBs, but the IC scattering is less efficient. 
Nevertheless, the recent detection
of the microquasar Cygnus~X-3 above 100~MeV by both \emph{Fermi}/LAT and
AGILE satellites \cite{Tavani2009,Abdo2009}, and the claim of short
flares from Cygnus~X-1 reported by \emph{AGILE} \cite{Sabatini2010}
confirmed the microquasars as interesting target for Very-High-Energy
(VHE) astrophysics. 

The MAGIC telescopes, with their low energy thresholds below 100 GeV,
are the best instrument among the current generation of imaging
atmospheric Cherenkov telescopes (IACTs) to search for VHE signals from
microquasars. 
MAGIC organized several observation campaigns of good
microquasar candidates such as Cygnus~X-3, Scorpius~X-1, SS~433, 
and Cygnus~X-1. These observations were carried out since 2006 and 
they will be described in detail in the next sections. 

\subsection{The MAGIC telescopes}
MAGIC consists of two 17~m diameter imaging atmospheric Cherenkov
telescopes (IACTs) located in the Canary island of La Palma, Spain
(2200~m above sea level). It became a stereoscopic system in Autumn
2009. The stereoscopic observation mode led to a significant
improvement in the instrument performance \cite{Carmona2011}. The
current sensitivity of the array yields 5$\sigma$ significance
detections above 250 GeV of fluxes as low as 0.8$\%$ of the Crab
Nebula flux in 50 hr. The energy threshold of the analysis lowered
down to 50 GeV for low zenith angle observations.

The data analysis was performed by using the standard MAGIC analysis 
software, which is described in detail in \cite{Carmona2011}.

\section{Cygnus X-3}
Cygnus X-3 is a microquasar \cite{Marti2001} consisting of an
accreting compact object of unknown nature and a Woft-Rayet companion
star \cite{vanKerkwijk1992}. This high-mass X-ray binary lies at a
distance of 7 kpc, and has an orbital period of 4.8 hours
\cite{Persignault1972}. 
Cygnus~X-3 displays two main X-ray spectral states, which are similar
to the canonical states of the black hole binaries \cite{Zdziarski2004}. The
hard state (HS) is dominated by a non-thermal power-law emission peaking at 20~keV \cite{Hjalmarsdotter2004}, whereas the soft state (SS) is characterized
by a strong thermal component with a non-thermal tail. A finer classification
of the spectral states was obtained by analyzing the correlation between
X-ray and radio fluxes \cite{Szostek2008}.\\

Cygnus X-3 was observed by the stand-alone MAGIC telescope  
for about 57 hours between March 2006 and August 2009. The results of
these observations are described in more detail at \cite{mine}, and
will only be summarized here. These VHE
observations where triggered by the source flux at other
wavelengths. In 2006, MAGIC received two alerts of a flaring state at
radio frequencies from the Russian RATAN-600 telescope, on March 10
and July 26, respectively. In 2007, a monitoring campaign of the
hard state of Cygnus X-3 was planned. The X-ray spectral state of the
source was defined by using public \emph{RXTE}/ASM (1.5--12~keV) and
\emph{Swift}/BAT (15--50~keV) data, as follows: a) \emph{Swift}/BAT
daily count rate larger than 0.05 counts cm$^{-2}$ s$^{-1}$, and b)
ratio between ASM and BAT count rates smaller than 200. During 2008
and 2009, MAGIC observed Cygnus~X-3 following the two high-energy (HE)
alerts issued by the \emph{AGILE} team on April 18, 2008 and July 18,
2009, respectively.\\

The search for a time-integrated VHE emission from Cygnus~X-3 was
performed by combining all the available data. It yielded no
significant excess events. The computed 95$\%$ confidence level (CL)
upper limit (UL) to the integral flux is of 2.2 $\times$ 10$^{-12}$
photons cm$^{-2}$ s$^{-1}$ for energies above 250 GeV. It corresponds
to 1.3$\%$ of the Crab Nebula flux at these energies.

\begin{figure}[!t]
\centering
\includegraphics[width=3.3in,height=10cm]{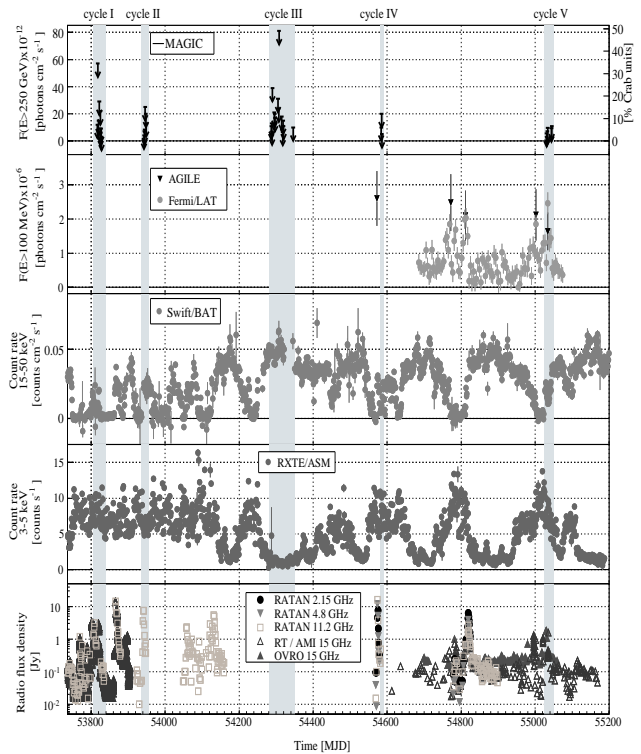}
\caption{Multiwalength behaviour of Cygnus X-3 from January 1, 2006
  until December 15, 2009. Refer to the text for the explanation.
Taken from \cite{mine}.}
\label{fig:CygX3_multi}
\end{figure}

\begin{figure}[!h]
\centering
\includegraphics[width=3.3in]{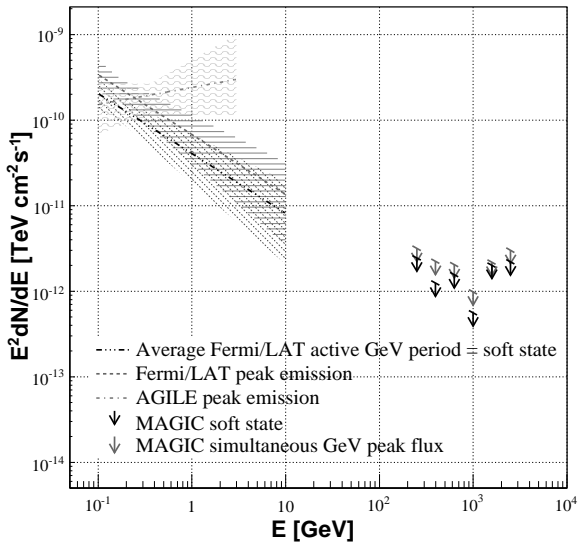}
\caption{Cygnus X-3 SED in the HE and VHE bands. The lines indicate
  the power-law spectra derived from \emph{Fermi}/LAT and \emph{AGILE}
  integral fluxes and photon indices. The arrows show the 95$\%$ CL
  MAGIC differential flux ULs and their slope indicates the assumed
  power-law spectrum with a photon index 2.6. Taken from \cite{mine}.}
\label{fig:CygX3_fermi}
\end{figure}

Since the source is variable on time scales of days at other
wavelengths, data from each day were analyzed separately. No signal was
detected in of the observation nights. These daily integral flux
ULs for energies above 250 GeV are shown in Figure
\ref{fig:CygX3_multi} together with HE $\gamma$-ray (\emph{AGILE} and
\emph{Fermi}/LAT $[$0.1--30 GeV$]$), hard X-ray (\emph{Swift}/BAT
$[$15--50~keV$]$), soft X-ray (\emph{RXTE}/ASM $[$1.5--12~keV$]$) and
radio measured fluxes from January 1, 2006 until December 15,
2009. This figure puts the MAGIC results in a multi-wavelength context
allowing us to derive the X-ray spectral state of the source during
MAGIC observations. \\
MAGIC pointed at Cygnus X-3 always when it was in its SS, except for the 2007
campaign when the HS was required by the observational trigger. The
two 2006 observation campaigns, triggered by radio flares, started
when the high-activity at radio frequencies had already ended, but
still the source was in a SS. Also in April 2008, following an
\emph{AGILE} alert, MAGIC started its observations of Cygnus~X-3 ten
days after a huge radio flare and an enhanced HE activity period, when the
source was still in the SS. The second \emph{AGILE} alert was much
more successful: MAGIC could take some data simultaneous with
a HE peak emission, detected by both \emph{AGILE} and \emph{Fermi}/LAT
\cite{Bulgarelli,Abdo2009}. The HE flux enhancement seems to occur
when the source is in its SS and is flaring at radio frequencies. The
$\sim$4 hours of simultaneous data showed no significant VHE signal,
yielding an integral flux at the level of 6$\%$ of the Crab Nebula
flux. \\
Figure \ref{fig:CygX3_fermi} shows the spectral energy distribution
(SED) of Cygnus X-3 above 100~MeV for both the SS and the HE peak
emission. The SED was computed by using the differential flux MAGIC
ULs, and the power-law spectra measured by both \emph{Fermi}/LAT and
\emph{AGILE} with a photon index of 2.7 and 1.8, respectively (refer
to \cite{mine} for further details). MAGIC ULs are compatible
with the extrapolation of the \emph{Fermi}/LAT power-law, but not with
the \emph{AGILE} extrapolation. The latter could suggest a spectrum cut-off
between some tens of GeV and 250 GeV. 
\section{Scorpius X-1}
Sco X-1, located at 2.8 $\pm$ 0.3 kpc, is a prototype Z-type low-mass
X-ray binary. It contains a neutron star orbiting around a M star
every 0.787 d. Z sources show a strong variability in the X-ray
intensity and colors in time scales of hours, and display a Z-shaped
track in the a hard color versus soft color diagram (CD). Sometimes,
the Z-track is better described by a double banana shape, such as for
Sco X-1. Along this track different X-ray spectral states can be
identified. They are known as Horizontal Branch (HB), Normal Branch
(NB) and Flaring Branch (FB) \cite{Hasinger1989}. Sco X-1 covers the
whole track in a few tens of hours, spending roughly half of the time in
the HB. Radio emission and a non-thermal power-law hard X-ray
emission have been only detected when the source is in the HB
\cite{DiSalvo2006}. These results indicate the presence of particle
accretion up to VHEs, and suggest a possible production of VHE
$\gamma$-rays via IC scattering in this spectral state.\\

\begin{figure}[!h]
\centering
\includegraphics[width=3.3in]{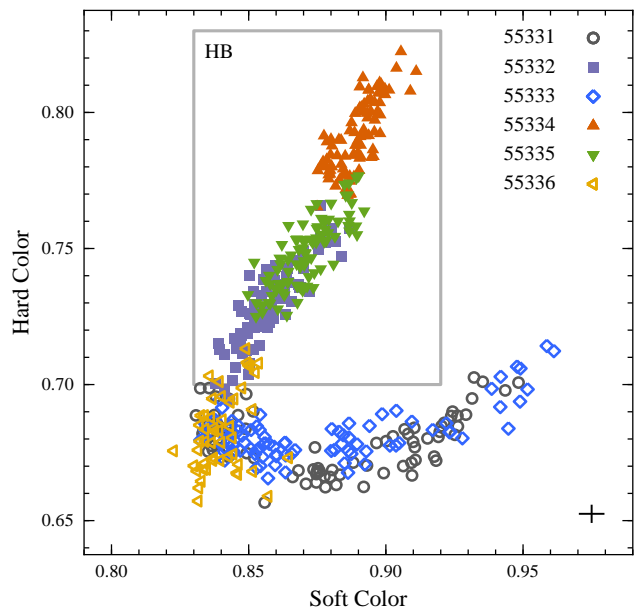}
\caption{X-ray color-color diagram of Sco X-1 from \emph{RXTE}/PCA
  data obtained during the simultaneous MAGIC campaign. Taken from \cite{Victor}.
\label{fig:ScoX1}}
\end{figure}

Simultaneous X-ray and VHE observations were carried out by MAGIC and
\emph{RXTE}/PCA in May 2010 for $\sim$8 hours. Results have been published 
in detail in \cite{Victor}. The X-ray data were
used to study the source X-ray spectral state during MAGIC
observations. The X-ray results are illustrated in Figure
\ref{fig:ScoX1} \cite{Victor}. During this campaign, the source
practically covered the full double-banana-shaped track. The
gray box in the figure indicates the data selected as HB. \\
The analysis of the MAGIC data showed no significant signal either in the
entire data sample, or in the HB state. Table \ref{tab:ScoX1}
shows the obtained ULs above 300 GeV. The ULs to the integral flux in
the HB state set an UL on the maximum TeV luminosity to jet power
ratio of 10$^{-3}$. This is a value similar to that obtained for Cygnus~X-3.
\cite{mine}.
\begin{table}[!t]
\centering
\begin{tabular}{cccc}
\hline
\hline
X-ray state & Effective time & \multicolumn{2}{c}{UL ($>$30 GeV)}\\
& (hr) & (cm$^{-2}$ s$^{-1}$) & C.U. \\
\hline
All & 7.75 & 2.4 $\times$10$^{-12}$ & 1.9$\%$\\
HB & 3.58 & 3.4 $\times$10$^{-12}$ & 2.7$\%$\\
NB/FB  & 4.17 & 2.8 $\times$10$^{-12}$ & 2.3$\%$\\
\hline
\end{tabular}
\caption{Integral flux ULs for the different X-ray spectral states of
  Sco X-1.
\label{tab:ScoX1}}
\end{table}

\section{Cygnus X-1}
Cygnus X-1 is the prime example of a black hole microquasar. It shows
the canonical X-ray HS and SS and a fast variability in different
wavelengths at distinct timescales. Radio emission stays stable during
the HS, and appears quenched during the SS \cite{Brocksopp1999}. Highly
collimated jets were detected in VLBA radio images suggesting that the
source is a microblazar \cite{Stirling2001,Romero2002}.\\

MAGIC monitored Cygnus X-1 for $\sim$ 50 hrs in 2006, and found an
evidence of signal at the level of 4.1$\sigma$ (post-trial) on
September 24th, 2006 \cite{Javi}. The source was in the HS and in
coincidence with a hard X-ray flare. Following this promising result, 
MAGIC observed the source for an additional 90 hrs between July 2007 and
November 2009. The last campaign was performed with the
stereoscopic system. However, no significant VHE signal was
found in this sample. 

\section{SS433}
SS433 is an eclipsing binary system containing a black hole orbiting
every 13.1 days around a supergiant star. The system displays
relativistic jets precessing with a period of 162.4 days in a cone of
half opening angle of 21$^\circ$ \cite{Margon1989}. The radio shell of 
W50 which surrounds the compact object is distorted by the jet 
precession. The latter extends in the east-west direction along the axis
of the jet precession and shows X-ray diffuse lobes symmetrically
displaced east and west of SS433 \cite{Safi1997}. These lobs, also
known as ``ears'', present a knotty structure.
The large jet and the diffuse non-thermal emission along the
jet make the microquasar into a good candidate for VHE $\gamma$-ray emission.
candidate. However, it is known that severe absorption effects
produced by the strong accretion disk wind could be at work for about
80$\%$ of the precessing period \cite{Reynoso2008}. 

MAGIC pointed towards SS 433 for 6 hrs in August 2008 and $\sim$10 hrs with
the stereoscopic system in May and June 2010. These observations were carried
out during precessional phases where the absorption of $\gamma$-rays
should be at its lowest. In addition, the eclipsing companion which
covers the jet inner region every $\sim$13 days was also avoided. 
MAGIC searched for VHE emission coming from both the central source
and the ``ears'', but no significant emission was found. The ULs to
the integral flux at energies above 150 GeV are at the level of 1$\%$
(3.6 $\times$ 10$^{-12}$ photons cm$^{-2}$ s$^{-1}$) and 5$\%$ (2.1
$\times$ 10$^{-11}$ photons cm$^{-2}$ s$^{-1}$) of the Crab flux for the
central source and the ``ears'' respectively. 

\section{Acknowledgments}
We would like to thank the Instituto de Astrofisica de 
Canarias for the excellent working conditions at the 
Observatorio del Roque de los Muchachos in La Palma. 
The support of the German BMBF and MPG, the Italian INFN,
the Swiss National Fund SNF, and the Spanish MICINN is gratefully acknowledged. 
This work was also supported by the Polish MNiSzW Grant N N203 390834, 
by the YIP of the Helmholtz Gemeinschaft, and by grant DO02-353
of the the Bulgarian National Science Fund. J.M. Paredes acknowledges financial support from ICREA Academia.\\

\clearpage

\end{document}